\documentclass[floatfix,twocolumn,preprintnumbers,amsmath,amssymb,superscriptaddress]{revtex4-2}
\usepackage{amsmath,amssymb,eucal,graphicx,float,epstopdf,xparse}
\usepackage{epsfig,subfigure}
\usepackage[utf8]{inputenc}
\usepackage{comment}

\def \equi#1{\mathrel{\mathop{\kern 0pt\sim}\limits_{#1}}} 
\usepackage{xcolor}

\usepackage[shortlabels]{enumitem}

\begin{document}
\title{Visitation Dynamics of $d$-Dimensional Fractional Brownian Motion}

\author{L\'eo R\'egnier}
\address{Laboratoire de Physique Th\'eorique de la Mati\`ere Condens\'ee,
CNRS/Sorbonne University, 4 Place Jussieu, 75005 Paris, France}
\author{Maxim Dolgushev}
\address{Laboratoire de Physique Th\'eorique de la Mati\`ere Condens\'ee,
CNRS/Sorbonne University, 4 Place Jussieu, 75005 Paris, France}
\author{Olivier B\'enichou}
\email{benichou@lptmc.jussieu.fr}
\address{Laboratoire de Physique Th\'eorique de la Mati\`ere Condens\'ee,
CNRS/Sorbonne University, 4 Place Jussieu, 75005 Paris, France}

\begin{abstract}
The fractional Brownian motion (fBm) is a paradigmatic strongly non-Markovian process with broad applications in various fields. Despite their importance, the properties of the territory covered by a $d$-dimensional fBm have remained elusive so far. Here, we study the visitation dynamics of the fBm
by considering the time  $\tau_n$ required to visit a site, defined as a unit cell of a $d$-dimensional lattice, when $n$ sites have been visited. Relying on scaling arguments, we determine all temporal regimes of the probability distribution function of $\tau_n$. These results are confirmed by extensive numerical simulations that employ large-deviation Monte Carlo algorithms. Besides these theoretical aspects, our results account for the tracking data of telomeres in the nucleus of mammalian cells, microspheres in an agorose gel, and vacuoles in the amoeba, which are experimental realizations of fBm.
\end{abstract}

\maketitle

The dynamics of many systems, such as observed for biological  \cite{Alessandro:2021,Krapf:2019} and other tracers in viscoelastic fluids \cite{Mason:1997,ernst2012fractional,Krapf:2019,dolgushev2024evidence}, show memory effects. These memory effects arise from interactions with the environment, leading to correlated displacements and anomalous diffusion. 
The fractional Brownian motion (fBm) is a paradigmatic model of such random motions with memory effects~\cite{Mandelbrot:1968}. Similarly to regular Brownian motion, fBm is a $d$-dimensional symmetric Gaussian process with stationary increments which are,  in contrast to the Brownian motion, correlated.    Explicitly, the process is defined by the covariance of the position $(x_1,\ldots,x_d)$ at times $t$ and $t'$:
\begin{align}
\text{Cov}[x_i(t),x_j(t')]&\equiv \left\langle x_i(t) x_j(t') \right\rangle-\left\langle x_i(t)\right\rangle \left\langle x_j(t') \right\rangle \nonumber  \\
&=\delta_{i,j}D(t^{2H}+t'^{2H}-|t-t'|^{2H}).\label{eq:cov}
\end{align}
Here, $0<H<1$ is the Hurst exponent and $D$ denotes the generalized diffusion constant \footnote{In the following, we work with dimensionless units and omit $D$. In numerical simulations, we take $D=1/8$ such that, at each unit time, the RW moves typically at a distance of $1/2$ lattice constant.}. Equation~\eqref{eq:cov} implies that for $H\neq1/2$ the process displays anomalous diffusion (subdiffusion for $H<1/2$ and superdiffusion for ${H>1/2}$), with a typical displacement growing as $t^H$.
The fBm has been shown to describe the subdiffusive motion of telomeres in the nucleus \cite{burnecki2012universal,Stadler:2017,Krapf:2019}, chromosomal loci~\cite{weber2010bacterial,bronshtein2015loss}, lipid granules in early mitotic cells \cite{jeon2011vivo}, beads in viscoelastic environments \cite{Mason:1997,ernst2012fractional,Krapf:2019,dolgushev2024evidence},  tracers in crowded fluids \cite{szymanski2009elucidating} and the superdiffusive motion of vacuoles inside an amoeba \cite{reverey2015superdiffusion,Krapf:2019}. 

\begin{figure}[t!]
    \centering
\includegraphics[width=\columnwidth]{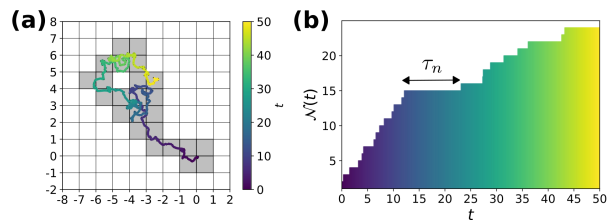}
    \caption{{\bf Visitation dynamics of fBm.} (a) A two-dimensional fBm trajectory ($H=0.75$) visiting unit cells of a square lattice (the cells visited are in gray). The temporal evolution of the trajectory is color-coded. (b) The number $\mathcal{N}(t)$ of cells visited as a function of time. $\tau_n$ is the time needed to visit a new cell after $n$ cells have already been  visited.}
    \label{fig:visit}
\end{figure}

Due to the non-Markovian nature of the fBm, obtaining  results for first-passage properties of this process is challenging \cite{van1992stochastic,Khadem:2021}. Nonetheless, several significant characteristics have been derived in one dimension, including the large time behavior of the survival probability of a target in unconfined space \cite{Krug:1997,Molchan:1999,levernier2019survival,Monch:2022,levernier2022everlasting}, the mean-first-passage time to a target in confinement \cite{Guerin:2016}, the statistics of record ages \cite{Regnier:2023b}, the fractal dimension of record sets \cite{Benigni:2018}, or the time at which the maximum is reached \cite{Delorme:2015}. In higher dimensions, the only available results concern first-exit times within wedges \cite{Qian:2003,Jeon:2014} and under confinement conditions \cite{Levernier:2018,Guerin:2016}. 

A paradigmatic observable of first-passage type, widely studied for Markovian random walks \cite{Hughes:1995},  is the extension of the domain explored by the process.
Despite their importance, for fBm, existing results on such exploration properties are limited to perturbative calculations around Brownian motion of the span \cite{Wiese:2019} and numerical simulations~\cite{Regnier:2023a}.  It is to be noted that, these results are restricted to dimension one. Finally, the exploration properties of fBm in dimensions larger than one have remained essentially unexplored so far.

Here, we characterize the exploration dynamics of  $d$-dimensional fBm based on the inter-visit times $\tau_n$ between visitations of new unit cells of a lattice (see Fig.~\ref{fig:visit} for description of the discretization procedure). The classical observable used to quantify the
exploration by a random walk (RW) on a lattice is the number $\mathcal{N}(t)$ of sites visited 
 at time $t$ \cite{Vineyard:1963}. However, being a cumulative quantity, $\mathcal{N}(t)$ discards important aspects of the exploration dynamics, e.g., it does not provide  information about the time needed to find a new site. To bridge this gap, the inter-visit time $\tau_n$, defined as the elapsed time between visits to the $n^\text{th}$ and $(n+1)^\text{th}$ distinct site, have been recently introduced in \cite{Regnier:2023a} and studied in the case of Markovian RWs. 

The statistics of $\tau_n$ is indeed crucial for foraging  dynamics \cite{Benichou:2016,Regnier:2024}, in which the living beings (such as bacteria \cite{Passino:2012} or animals \cite{orlando2020power}) cannot remain too long  without nutrition. 
Other important examples of situations controlled by these inter-visit times include the trapping of diffusing molecules \cite{hollander_weiss_1994}, the spread of successful strategies in populations according to evolutionary game theory \cite{traulsen2009exploration}, the construction of multidimensional equivalence classes in deep neural networks \cite{Benfenati:2024}, visitation dynamics on networks \cite{Benatti:2023}, and space exploration by robots \cite{Lepowsky:2024}, to name a few. 

Determining the statistical properties of $\tau_n$ for fBm in dimension $d>1$ is challenging, because domains of $n$ sites visited display a variety of shapes: The territory visited is typically non-spherical, contains holes and islands, expands constantly, and depends on the entire previous history of the walker. Here, relying on scaling arguments, we show that for a fBm with Hurst exponent $H$, in dimension $d$, the probability distribution function (pdf) of $\tau_n$, $F_n(\tau)=\mathbb{P}(\tau_n=\tau)$, is entirely characterized by the single exponent $\mu\equiv dH$. This parameter defines the nature of the exploration \cite{Hughes:1995}: recurrent ($\mu< 1$) and marginally recurrent ($\mu=1$), where all sites are eventually visited, or transient ($\mu>1$), where some sites are never visited. We go beyond the results of Ref.~\cite{Regnier:2023a} obtained for Markovian RWs, and show that there are  (i) an early-time regime $\tau\ll \vartheta_n$ (with $\vartheta_n=n^{1/\mu}$ for recurrent fBm,  $\vartheta_n=n^{1/2}$ for marginal fBm, and $\vartheta_n=1$ for transient fBm) characterized by algebraic decay, $F_n(\tau)\propto 1/\tau^{1+\mu}$, (ii) an intermediate regime $\vartheta_n\ll \tau\ll \Theta_n$ ($\Theta_n=n^{1/\mu}$ for recurrent fBm, $\Theta_n=n^{3/2}$ for marginal fBm, and $\Theta_n=n^{1+1/\mu}$ for transient fBm) where the statistics exhibit stretched-exponential decay, $F_n(\tau)\propto \exp\left[-\left(\tau/\vartheta_n \right)^{\mu/(1+\mu)} \right]$, and finally (iii) an exponential decay at large times (for all fBm types). Note that the behaviour of characteristic times $\vartheta_n$ and $\Theta_n$ is such that all three regimes are observed for the marginal fBm only. These analytical results are confirmed by numerical simulations.
Importantly, we demonstrate that our theoretical results describe the exploration dynamics of various biological tracers,  known to be  experimental realizations of fBm\cite{Krapf:2019}.

\textit{Early-time regime.} We first focus on the early-time regime of  recurrent and marginally recurrent fBm ($\mu=Hd\leq 1$). In this regime, the region of $n\gg1$ sites already visited  appears as  effectively infinite.   We thus expect an algebraic decay    
\begin{align}
   F_n(\tau) \propto 1/n^{\epsilon}\tau^{1+y} \label{eq:algebraic}
\end{align}
with a potential dependence on $n$  involved only in the prefactor of $F_n(\tau)$. In order to find the exponents $y$ and $\epsilon$, we need to go beyond the approach used in Ref.~\cite{Regnier:2023a} for Markovian processes, which explicitly relies on a  renewal type equation, which does not hold for a non-Markovian process like the fBm.  We thus develop a  scaling approach, which describes the visitation dynamics for general scale-invariant non-Markovian processes $x(t)$ with stationary increments, satisfying $\langle x(t)^2\rangle \propto t^{2/d_\text{w}}$, where  $d_\text{w}$ is the walk dimension ($d_\text{w}=1/H$ for fBm) (see SM \footnote{See Supplemental Material at XXX} for details and an alternative derivation of the exponents $y$ and $\epsilon$.).

\begin{figure}[t!]
    \centering
\includegraphics[width=\columnwidth]{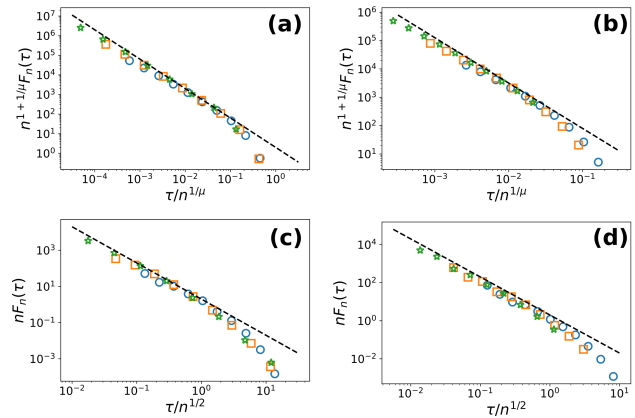}
    \caption{{\bf Early time statistics.} The rescaled distributions of the $n^\text{th}$ intervisit time for fBm in (a) $d=2$, $H=0.25$ ($n=160,$ $320$ and $640$; recurrent fBm with $\mu=0.5$), (b) $d=3$, $H=0.2$ ($n=160,$ $320$ and $640$; recurrent fBm with $\mu=0.6$) (c) $d=3$,  $H=1/3$ ($n=10^2$, $10^3$ and $10^4$; marginal fBm with $\mu=1$) (d) $d=4$, $H=1/4$ ($n=10^2$, $10^3$ and $10^4$; marginal fBm with $\mu=1$). The black dashed lines stand for the algebraic decay of Eq.~\eqref{eq:earlyFBM}.}
    \label{fig:early}
\end{figure}

The first step consists in determining  the scalings of the time ${T_n=\sum_{k=0}^{n-1} \tau_k}$ to visit $n$ sites and of its increment $T_{n+m}-T_n$ by relying on the scale-invariance property of $x(t)$. For recurrent walks, the number of  sites visited is proportional to the volume spanned by the walker, of linear extension $ x(t) \propto t^{1/d_\text{w}}$, so that the number of sites visited corresponds to $\mathcal{N}(t)\propto \langle x(t)^2\rangle^{d/2} \propto t^{d/d_\text{w}}$  \cite{Meroz:2013}. Finally, $\mathcal{N}(T_n)=n\propto T_n^{d/d_\text{w}}$, i.e. $T_n \propto n^{d_\text{w}/d}$. 
Next, we note that at time $T_n$, the fBm just visited a new site and is thus at the boundary of the visited domain. 
Hence the number of new sites visited during the time interval $[T_n,T_n+t]$ (with $t\ll T_n$)
correspond to a fraction of the volume of the ball of radius $x(T_n+t)-x(T_n)$ (see SM for numerical check):
\begin{align}
\hspace{-0.9mm}    \mathcal{N}(T_n+t)-\mathcal{N}(T_n)\propto \left( x(T_n+t)-x(T_n) \right)^{d}\propto t^{d/d_\text{w}}. \label{eq:scalingNTt}
\end{align}
Here, we used the stationarity of the increments, so that the distribution of $x(T_n+t)-x(T_n)$ is independent of $n$. 
Introducing $\mathcal{N}(T_{n+m})-\mathcal{N}(T_n)=m$ and  $t=T_{n+m}-T_n$, we finally obtain from Eq.~\eqref{eq:scalingNTt} that, for $1\ll m \ll n$,
\begin{align}
    T_{n+m}-T_n\propto m^{d_\text{w}/d} \; .\label{eq:scaling_incrT}
\end{align}
\begin{figure}[ht!]
    \centering
\includegraphics[width=0.95\columnwidth]{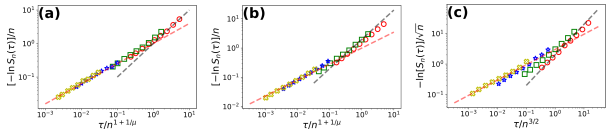}
    \caption{{\bf Intermediate and long time regimes.} Rescaled tail distribution of the intervisit time, $S_n(\tau)\equiv \int_\tau^\infty F_n(\tau'){\rm d}\tau'$, as a function of $\tau/\Theta_n$ for (a) fBm in $2d$ of $H=0.75$ (for $n=10$, $20$, $80$ and $160$; transient fBm with $\mu=1.5$) (b) fBm in $3d$ of $H=0.4$ ($n=10$, $20$, $80$ and $160$; transient fBm with $\mu=1.2$) (c) fBm in $3d$ of $H=0.33$ ($n=10$, $25$, $100$ and $400$; marginal fBm with $\mu\approx1$), where the red line is proportional to $\left(\tau/\Theta_n\right)^{\frac{\mu}{1+\mu}}$ and the grey line is linear. Increasing values of $n$ are represented in red circles, green squares, blue stars and yellow crosses.   }
    \label{fig:stretched}
\end{figure}

The second step relies on an alternative derivation of the scaling behavior of the increments $ T_{n+m}-T_n$, by connecting them to the inter-visit times $\tau_k$. We use that a sum of broadly distributed random variables (with  infinite mean) is dominated by the largest term of the sum \cite{bouchaud1990anomalous,Vezzani:2019}, i.e.,
\begin{align} \label{eq:max_indep}
    T_{n+m}-T_n=\sum_{k=n}^{m+n-1}\tau_k\approx \max \lbrace \tau_k|n\leq k<n+m \rbrace,
\end{align}
where we assume, as self-consistently checked below, that $0<y\leq 1$ in Eq.~\eqref{eq:algebraic} and that the random variables $\tau_k$ are effectively independent. Using the effective independence of the correlations between the $\tau_k$, we obtain that (for $1 \ll m \ll n$):
\begin{align} \label{eq:indep_distrib}
    \mathbb{P}(T_{n+m}-T_n \leq T)&\approx \prod_{k={n}}^{n+m-1}\left(1-\int_{T}^\infty F_k(\tau){\rm d}\tau \right)\\
    &\approx \exp\left[-\mathrm{ const.}\frac{m}{n^\epsilon T^y} \right].
\end{align}
In particular, this implies that $T_{n+m}-T_n\propto m^{1/y}n^{-\epsilon/y}$. 

The last step consists in comparing this last result to Eq.~\eqref{eq:scaling_incrT}, which   leads to  $\epsilon=0$ and $y=d/d_\text{w}\leq 1$. Finally, for a recurrent fBm in a medium of dimension $d$, we obtain the behavior of the inter-visit time statistics at early times (see Fig.~\ref{fig:early} for numerical check and discussion below):
\begin{align}
    F_n(\tau)\propto \frac{1}{\tau^{1+dH}}. \label{eq:earlyFBM}
\end{align}
This algebraic regime holds as long as the region  of sites 
visited appears as effectively infinite. This leads to the definition of  the crossover time $\vartheta_n$ as the escape time from the largest domain fully visited,  when $n$  sites have been visited. It can be shown (see SM) that this domain  typically contains  $n$ sites for $\mu<1$, $n^{1/2}$ sites for $\mu=1$ and a constant number of sites for $\mu>1$. Hence, $\vartheta_n=n^{d_\text{w}/d}$ for $\mu<1$, $\vartheta_n=\sqrt{n}$ for $\mu=1$, and $\vartheta_n=1$ for $\mu>1$.

We now self-consistently check the effective independence of the $\{ \tau_k \}$ employed in Eq.~\eqref{eq:max_indep}. This builds upon the argument presented originally in Ref.~\cite{Carpentier:2001} and extended in Refs.~\cite{Regnier:2023b,Regnier:2024}. While the argument was initially formulated for Gaussian correlated but identically distributed random variables, we apply it here to non-Gaussian and non-identically distributed random variables. The key point is that the effect of correlations on the statistics of the maximum can be ignored if these correlations are negligible with respect to the fluctuations of the maximum in the absence of  correlations. Here, the typical correlation between the $\lbrace \tau_k \rbrace$ can be estimated as $\text{Cov}\left[\tau_{n+m/4},\tau_{n+3m/4} \right]$, which is decaying with $m$: the distance between the two sites at which the fBm starts the visitation (after visits of $n+m/4$ and $n+3m/4$ sites) increases with $m$, and correlations between the corresponding trajectories decay algebraically with time and thus with the number of  sites visited. Regarding the maximum, Eqs.~\eqref{eq:scaling_incrT} and \eqref{eq:max_indep} show that its fluctuations without correlations are typically given by $m^{d_\text{w}/d}$ using Eqs.~\eqref{eq:scaling_incrT} and \eqref{eq:max_indep}. This indicates that, when $m$ is large, one can neglect the correlations between the $\tau_k$ (see also SM for details and numerical check), which justifies both Eqs. \eqref{eq:max_indep} and \eqref{eq:indep_distrib}.

As an important result, Eq.~\eqref{eq:earlyFBM} stands in sharp contrast with the search problem of a single target, for which the first-passage time is also algebraically distributed, $F(\tau)\propto \tau^{-1-\theta}$, but  where  the persistence exponent $\theta$ is given by $\theta= 1-dH $ \cite{Levernier:2018}. This reflects the qualitative difference between   the exit time statistics from the complex random domain  considered here and the first-passage time to a single target.

\textit{Intermediate and long time regimes.} We now turn to the intermediate and long time regimes. We follow the arguments used in \cite{Regnier:2023a} for Markovian RWs, which in fact are  general and hold also for non-Markovian processes.
The key idea, in analogy with the trapping problem \cite{Donsker:1979}, is that in the intermediate time regime, the statistics is dominated by realizations containing a large region free of "traps" (the non-visited sites). We denote by $Q_n(r)$ the distribution  of the radius $r$ of the largest ball fully visited. We show in SM that  (i) the typical radius $\rho_n$  of this largest ball fully visited   grows as $n^{1/d}$ for $\mu<1$, $n^{1/2d}$ for $\mu=1$ and more slowly than any power of $n$ for $\mu>1$ and (ii) $Q_n(r)$ decays exponentially as a function of  $(r/\rho_n)^d$.
Then, we consider the probability $S_n(\tau)\equiv \int_\tau^\infty F_n(\tau'){\rm d}\tau'$ to escape the domain visited 
after time $\tau$. A lower bound for this quantity is provided by replacing the domain visited  by the  largest ball fully visited in the domain. Additionally, we use that the probability to remain inside a spherical region of radius $r$ up to a time $\tau$ is given by $\exp\left(-\tau/r^{1/H} \right)$ (prefactors independent of $\tau$ and $r$ are put to one for simplicity). Then, by summing over all possible  values of the radius up to  $r=n^{1/d}$, we have $S_n(\tau)\geq \int^{n^{1/d}}_0 {\rm d}r Q_n(r) e^{-\tau/r^{1/H}} $. Using next the expression of $Q_n(r)\propto \exp\left[-\left(r/\rho_n\right)^d\right]$ and a  saddle point method, we get for the lower bound of the probability $S_n(\tau)$,
\begin{align}
    S_n(\tau)&\approx  \int^{n^{1/d}}_0 {\rm d}r  e^{-(r/\rho_n)^d-\tau/r^{1/H}}
    \propto e^{-\left( \tau/\vartheta_n\right)^{\mu/(1+\mu)}} \; . \label{eq:stretched}
\end{align}
Similarly to the classical trapping problem \cite{hollander_weiss_1994,Hughes:1995,Donsker:1979}, we have used in \eqref{eq:stretched}  the fact that the lower bound is expected to actually provide the scaling behavior of $S_n(\tau)$, as we numerically check in Fig.~\ref{fig:stretched} and discuss below. \footnote{Note that, in Eq.~\eqref{eq:stretched}, we omit algebraic prefactors in $\tau$ and $n$ which are not accounted for by our scaling approach.} The stretched exponential regime of Eq.~\eqref{eq:stretched} breaks down when the radius $r^*(\tau)$, which minimizes $U_\tau(r)=\tau/r^{1/H}+(r/\rho_n)^d$, reaches the maximal value $n^{1/d}$. This defines the time $\Theta_n$ after which the decay is purely exponential, $r^*(\Theta_n)=n^{1/d}$, such that $\Theta_n=n^{3/2}$ for $\mu=1$ and $\Theta_n=n^{1+1/\mu}$ for $\mu>1$. Note that, for recurrent fBm ($\mu<1$), the time $\Theta_n$ is of the same order as the crossover time $\vartheta_n$, so that only the exponential decay is observed at long times. For all values of $\mu$, the time scale associated with the exponential decay is given by $n^{1/\mu}$, which corresponds to the typical exit time of the ball of volume $n$. 

Finally,  Eqs.~\eqref{eq:earlyFBM} and \eqref{eq:stretched} provide  all the  asymptotic regimes of the inter-visit statistics of a $d$-dimensional fBm.

\textit{Numerical check.} 
Next, we move on to the numerical check of Eqs.~\eqref{eq:earlyFBM} and \eqref{eq:stretched}. 
Concerning the algebraic regime, we rely on the standard Davies-Harte algorithm \cite{Davies:1987} which generates fBm trajectories of arbitrary Hurst index. By considering $d$ such independent trajectories and taking the lattice discretization as in Fig.~\ref{fig:visit}, we obtain the inter-visit time statististics presented in Fig.~\ref{fig:early}, which unambiguously confirms the expected algebraic decay of exponent $1+\mu$  of Eq.~\eqref{eq:earlyFBM} in both the recurrent and marginal cases.  
 
To get access to the large deviation regime of Eq.~\eqref{eq:stretched}, it is necessary to go beyond the above-described method and to adapt the importance sampling approach introduced in Refs.~\cite{Hartmann:2013,Hartmann:2024}. The general idea of the method  (see SM for details) is to use Monte-Carlo Markov chain techniques, to bias the fBm trajectories towards rare inter-visit times realizations. Then, we reweight the biased distribution to obtain the (unbiased) tail distribution $S_n(\tau)$ of the inter-visit time statistics presented in Fig.~\ref{fig:stretched}..

\begin{figure}[t!]
    \centering    \includegraphics[width=\columnwidth]{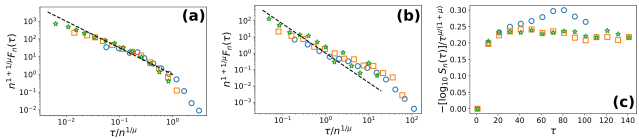}
    \caption{{\bf Inter-visit time $\tau_n$ statistics of different tracers in various systems of dimension $d=2$.} {\bf (a)} Telomere in the nucleus of mammalian cells ($\mu=dH=0.5$, recurrent). {\bf (b)} Microsphere in agarose gel ($\mu=0.8$, recurrent). {\bf (c)} Vacuole in amoeba ($\mu=1.3$, transient).   Blue circles, orange squares and green stars represent $\tau_n$ for $n=5$, $10$, $15$ sites visited, respectively; black dashed lines stand for $1/x^{1+\mu}$ behavior with $x=\tau_n/n^{1/\mu}$.}
    \label{fig:tracer}
\end{figure}

For both the marginal and transient cases, we observe that the properly rescaled tail distribution $S_n(\tau)$ exhibits a stretched exponential regime with the exponent $\mu/(1+\mu)$, as indicated by the red lines in Fig.~\ref{fig:stretched}. Indeed, for times $\vartheta_n\ll \tau\ll \Theta_n$, from Eq.~\eqref{eq:stretched},  $[-\ln S_n(\tau)]/n\sim \tau^{\frac{\mu}{1+\mu}}/n=\left(\tau/\Theta_n \right)^{\frac{\mu}{1+\mu}}$ (transient case) and $[-\ln S_n(\tau)]/\sqrt{n}\sim \sqrt{\tau/(n\sqrt{n})}=\sqrt{\tau/\Theta_n}$ (marginal case). This stretched exponential regime breaks down at the time scales predicted above: $\Theta_n = n^{3/2}$ for the marginal case and $\Theta_n = n^{1+1/\mu}$ for the transient case. We also confirm the exponential decay of $S_n(\tau)$ at times $\tau\gg \Theta_n$, i.e. $[-\ln S_n(\tau)]/n\sim \tau/n^{1+1/\mu}=\tau/\Theta_n$ in the transient case and $[-\ln S_n(\tau)]/\sqrt{n}\sim \tau/n^{3/2}=\tau/\Theta_n$ in the marginal case, indicated by the grey lines in Fig.~\ref{fig:stretched}. Note that, due to the existence of two non-trivial time scales in the marginal case ($\vartheta_n$ and $\Theta_n$, both of which grow with $n$), there is no collapse of the curves at small $\tau$, corresponding to the algebraic early-time regime. 

Finally, Fig.~\ref{fig:stretched} unambiguously confirms the asymptotic regime given by Eq.~\eqref{eq:stretched}.

\textit{Application: Time between visits of  new sites by tracers in $2d$ environments.} We now apply our results to the data from Refs.~\cite{Krapf:2019,Stadler:2017,reverey2015superdiffusion}: The temporal trajectories of a telomere in the nucleus of mammalian cells ($H=0.25$) \cite{Stadler:2017,Krapf:2019} and of microspheres in an agarose gel ($H=0.4$)  in $2d$ \cite{Krapf:2019}, both having a recurrent fBm behaviour ($\mu=0.5$ and $0.8$, respectively); the motion of vacuoles in amoeba ($H=0.65$, $d=2$) \cite{reverey2015superdiffusion,Krapf:2019}, which is transient ($\mu=1.3$).
We stress that all these systems are well-described by fBm \cite{Krapf:2019}, so that our non-Markovian framework is required to describe their exploration properties (see  SM for detailed analysis). Parts (a) and (b) of Fig.~\ref{fig:tracer} corresponding to the telomeres' and microspheres' trajectories are well-described by the algebraic regime of Eq.~\eqref{eq:earlyFBM}. Part (c) of Fig.~\ref{fig:tracer} corresponding to  the motion of vacuoles in amoeba verifies the stretched-exponential regime of Eq.~\eqref{eq:stretched} (as $[-\log S_n(\tau)]/\tau^{\mu/(1+\mu)}$ is constant at sufficiently large $n$ and $\tau$). This shows that our formalism accounts for the exploration dynamics of experimentally relevant non-Markovian tracers.

{\it Conclusion.} In this letter, we have studied the exploration dynamics of $d$-dimensional fBm's of Hurst exponent $H$ by considering the time $\tau_n$ elapsed between the discoveries of two successive new sites. We have shown that the scaling behavior of $\tau_n$ statistics is completely characterized by the exponent $\mu=dH$. For $\mu<1$ (recurrent case), the pdf of $\tau_n$ has an algebraic decay (Eq.~\eqref{eq:earlyFBM}) followed by an exponential one; for $\mu>1$ it has a stretched exponential decay (Eq.~\eqref{eq:stretched}) and an exponential one; the marginal case $\mu=1$ combines all types of decays (algebraic, stretched-exponential, and exponential). These results, confirmed by extensive simulations, account for experimental data of a variety of non-Markovian tracers.

\begin{acknowledgements}
 We are thankful to D. Krapf, M. Weiss, F. Taheri and C. Selhuber-Unkel for providing us the experimental trajectories of fBm realizations used in Ref. \cite{Krapf:2019}. We thank J. Br\'emont and  P. Viot for useful discussions.  
\end{acknowledgements}
\newpage

\bibliographystyle{apsrev4-1}

\end{document}


\RestyleAlgo{ruled}
	\title{SUPPLEMENTAL MATERIAL\\
		Visitation Dynamics of $d$-Dimensional Fractional Brownian Motion}
	
	\author{L. R\'egnier}
	\author{M. Dolgushev}
	\author{O. B\'enichou}
	\maketitle
	
	\tableofcontents
	\section{Properties of the set of visited sites}
	\label{sec:Prop}
	
	In this section, we focus on the visitation properties of $d$-dimensional fractional Brownian motion (fBm). Specifically, we consider the following properties of the set of visited sites: the scaling with time of the average number ${N(t)=\left\langle \mathcal{N}(t)\right\rangle}$ of visited sites (defined as unit cells of a $d$-dimensional hypercubic lattice), the scale invariance of both the number of visited sites $\mathcal{N}(t)$ and the time $T_n$ to visit $n$ sites, as well as the scaling with $n$ of the radius $\rho_n$ of the largest fully visited ball when $n$ sites have been visited. While all these properties were studied for regular Markovian random walks, to the best of our knowledge, they are not known for $d$-dimensional fBm.
	
	\subsection{Average number of visited sites} \label{sec:Nt_av}
	
\begin{figure}
		\centering
		\includegraphics[width=\columnwidth]{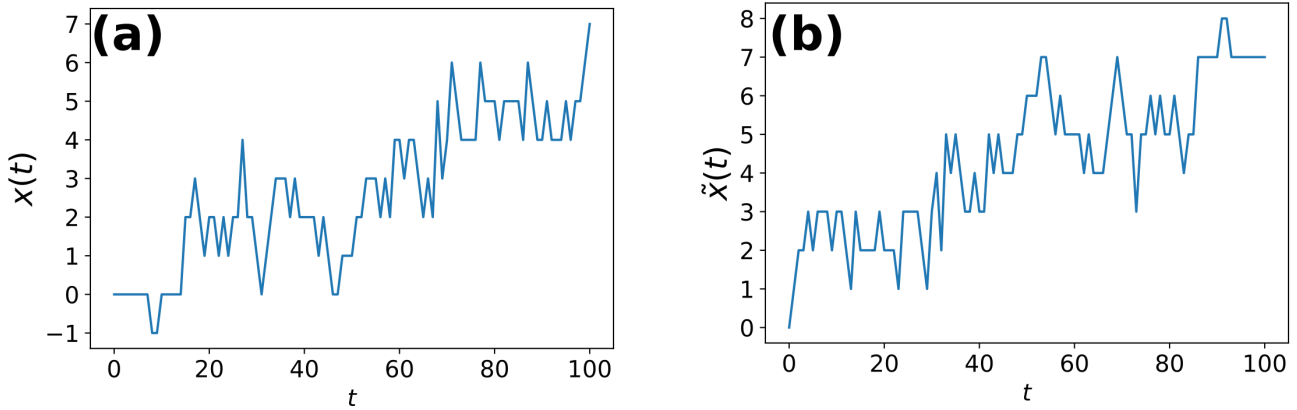}
		\caption{{\bf Mapping between trajectories of visiting new sites at $t$ and of not returning at $0$ by $t$.} {\bf (a)} A discretized trajectory $x(t)$ of an fBm of $H=0.25$ contributing to the probability $\Delta_t$ of visiting a new site at time $t$ (here position $x=7$). {\bf (b)} Time reversed trajectory $\tilde{x}(t)$, which does not return to $0$ by time $t$.}
		\label{fig:Map}
	\end{figure}

	First, the average of the number $\mathcal{N}(t)$ of sites visited by time $t$ grows as 
	\begin{align} \label{eq:av_N}
		N(t)\equiv \left\langle \mathcal{N}(t)\right\rangle\sim \begin{cases}
			t^{\mu} \; \text{(recurrent)}\\
			t \; \text{(marginal, up to log-corrections)}\\
			t \; \text{(transient)}
		\end{cases}
	\end{align}
	where $\mu\equiv H d$, $H$ being the Hurst exponent and $d$ the spatial dimension. Inspired by the Dvoretzky-Erd\H{o}s lemma \cite{Hughes:1995} for homogeneous Markovian walks, by noting $\Delta_{t'}$ the indicator function of a new visit at time $t'$, $N(t)$ can be viewed as the sum of the probabilities $\left\langle \Delta_{t'} \right\rangle$ to visit a new site at step $t' \leq t$, 
	\begin{align} \label{eq:Nt_sum}
		N(t)=\sum_{t'\leq t} \left\langle\Delta_{t'}\right\rangle \; .
	\end{align}
	By time reversal invariance of the fBm, $\left\langle\Delta_{t'}\right\rangle$ is exactly the probability of never returning to the origin up to $t'$, see \ref{fig:Map} for a $1d$ example (but which is valid in any dimension). If the trajectory $\left( x(t) \right)_{0\leq t \leq T}$ is associated to the increments $\eta_{t}=x(t+1)-x(t)$, then the time reversed trajectory at time $T$ defined for any $t$ 
	by $\tilde{x}(t)=\sum_{t'=0}^{t-1}\eta_{T-t'}$ has the same probability as $x(t)$. Indeed, it is still Gaussian with the same covariance function given by
	\begin{align}
		\text{Cov}\left[\tilde{x}(t), \tilde{x}(t')\right]&=\sum_{i\leq t,j\leq t'}  \text{Cov}\left[\eta_{T-i}, \eta_{T-j}\right]\\
		&=\sum_{i\leq t,j\leq t'}  \text{Cov}\left[\eta_{i}, \eta_{j}\right]\\
		&=\text{Cov}\left[x(t), x(t')\right]
	\end{align}
	where we used that the $\eta_{t'}$ are invariant by translation and invariant by time reversal,
	\begin{align}
		\text{Cov}\left[\eta_{T-i}, \eta_{T-j}\right]= D\left( |j-i+1|^{2H}+|j-i-1|^{2H}-2|i-j|^{2H}\right)=\text{Cov}\left[\eta_{i}, \eta_{j}\right] \; .
	\end{align}
	From this, we observe that visiting a new site at time $t$ for the  trajectory $x$ corresponds exactly to not having returned to the origin for the time-reversed trajectory $\tilde{x}$ up to time $t$ (see \ref{fig:Map}). This is why $\left\langle\Delta_{t}\right\rangle$ corresponds exactly to the probability not to return to the origin by time $t$.
	
	In the transient case $\mu>1$, $\left\langle\Delta_{t'}\right\rangle$ converges at large $t'$ to a finite value corresponding to the probability to never return to the origin, we note $1-\mathcal{R}$ (the RW is transient, see Ref.~\cite{Levernier:2018}). This implies that for $\mu>1$, at large times,
	\begin{align}
		N(t)=\sum_{t'\leq t} \left\langle\Delta_{t'}\right\rangle \sim (1-\mathcal{R})t  \; .
	\end{align}
	
\begin{figure}[th!]
		\centering
		\includegraphics[width=\columnwidth]{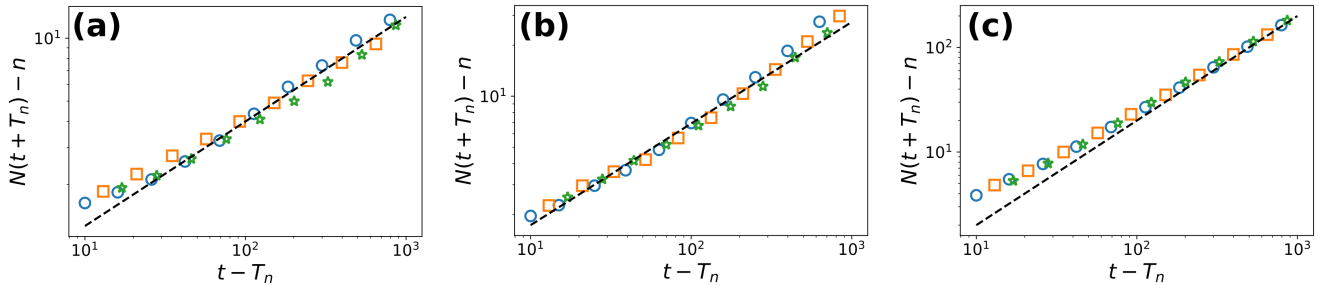}
		\caption{{\bf Aged visitation.} We compute the average number of additional sites visited after having visited $n$ sites, $N(t+T_n)-n$, that we compare to the $t^\mu$ growth ($\mu=dH$, dashed lines) for (a) $2d$ fBm of $H=0.25$ (recurrent; $n=100$, $200$ and $400$ in blue circles, orange squares and green stars); (b) $3d$ fBm of $H=0.2$ (recurrent; $n=100$, $200$ and $400$ in blue circles, orange squares and green stars); (c) $3d$ fBm of $H=1/3$ (marginal; $n=1000$, $2000$ and $4000$ in blue circles, orange squares and green stars).}
		\label{fig:Nt_Tn}
	\end{figure}

	For recurrent fBm $\mu<1$, we use the results of Refs.~\cite{Molchan:1999,Krug:1997,levernier2019survival}   which give the probability not to have returned to the origin by time $t'$ (the probability $\left\langle \Delta_{t'} \right\rangle $) that decays as
	\begin{align} \label{eq:Delta_rec}
		\left\langle \Delta_{t'} \right\rangle \propto 1/t'^{1-\mu} \; ,
	\end{align}
	which results in 
	\begin{align}
		N(t)=\sum_{t'\leq t} \left\langle \Delta_{t'} \right\rangle \propto t^\mu \; .
	\end{align}
	The growth of $N(t)$ with $t$ is solely characterized by the recurrence (eventually marginal) or transience properties of the process. We check in \ref{fig:Nt_Tn} that this growth is the same even when the walker has already visited $n$ sites, the result we use in Eq.~(3) of the main text.

\begin{figure}[th!]
		\centering
		\includegraphics[width=0.7\columnwidth]{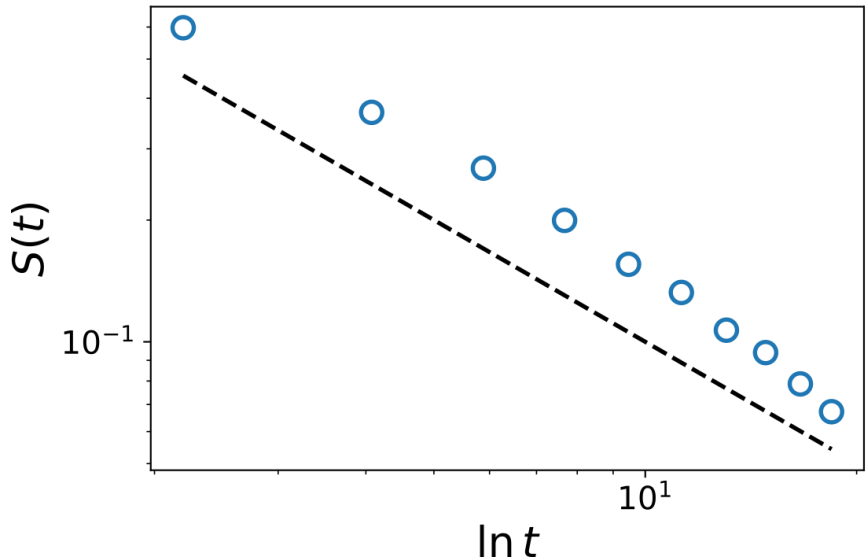}
		\caption{{\bf First return time to the origin of marginal fBm.} Tail distribution $S(t)$ of the first return time to the origin  (blue circles) for a marginal fBm of Hurst exponent $H=1/d$ in dimension $d=3$. The black dashed line corresponds to $1/\ln t$.}
		\label{fig:St_marg}
	\end{figure}

	For marginal fBm with $\mu=1$, the probability $\left\langle \Delta_{t'} \right\rangle$ not to have returned to the origin by time $t'$ decays slower than algebraically, in agreement with the decrease with time in Eq.~\eqref{eq:Delta_rec}. To show this, we develop an argument similar to the one proposed in Ref.~\cite{Krug:1997} that provides a logarithmic decay of probability of the first return time to the origin. The idea is the following: let us note  $\mathcal{N}_{x(i)}(t)$ the number of visits of site $x(i)$ during the time interval $[i,i+t]$. Because the fBm is invariant by time translation, the distribution of $\mathcal{N}_{x(i)}(t)$ is the same as the distribution of $\mathcal{N}_{0}(t)$, the number of returns to the origin during the time interval $[0,t]$. Besides, the average $\mathcal{N}_{0}(t)$ grows at a rate given by the probability to be at the origin at time $t$, which is given by (using that the fBm is a Gaussian process of standard deviation $t^{H}$) 
	\begin{align}
		\frac{{\rm d} \left\langle \mathcal{N}_{0}(t) \right\rangle }{{\rm d}t }= \frac{1}{\left( \sqrt{4 \pi D t^{2H}}\right)^d} \propto 1/t \; .
	\end{align}
	Consequently, we have a logarithmic growth of the average number of returns to the origin during the time interval $[0,t]$, 
	\begin{align} \label{eq:return_origin}
		\left\langle \mathcal{N}_{0}(t) \right\rangle \propto \ln t \; .
	\end{align}
	Then, we note that the number $t$ of steps made by the fBm must be equal to the total number of returns to each visited site (which is logarithmic for each of these sites according to Eq.~\eqref{eq:return_origin}). In other words, we have a mean-field type equation which connects the average number $N(t)$ of visited sites to the average number of return to these sites,
	\begin{align}
		N(t) \left\langle \mathcal{N}_{0}(t) \right\rangle \approx t
	\end{align}
	which results in Eq.~\eqref{eq:marg_Nt}. More precisely, the exact expression is given by first considering all times $t'$ where a new visit is made (indicated by $\Delta_{t'}$) and then counting  the number of returns to these sites :
	\begin{align}
		\sum_{t'=0}^t \Delta_{t'} \mathcal{N}_{x(t')}(t-t') =t \; .
	\end{align}
	Taking the average and neglecting the correlation between the first visitation event and the number of returns, we obtain the following equation
	\begin{align} \label{eq:exact_marg}
		\sum_{t'=0}^t \left\langle \Delta_{t'} \right\rangle \left\langle \mathcal{N}_{0}(t-t')  \right\rangle =t \; .
	\end{align}
	Entering the asymptotics of Eq.~\eqref{eq:return_origin} into Eq.~\eqref{eq:exact_marg}, we obtain that 
	\begin{align} \label{eq:St_marg}
		\left\langle \Delta_{t} \right\rangle\propto \frac{1}{\ln t} \; .
	\end{align}
	Eventually a semi-Markovian approximation \cite{Jeon:2014} also states that  $\left\langle \Delta_{t} \right\rangle$ decays as $1/\ln t$ for marginal fBm (so that, for $Hd=1$, the process is marginally recurrent in the sense defined in Ref.~\cite{Levernier:2018}). However, our argument uses a  weaker assumption by neglecting correlations between new visit events and the number of returns. Nevertheless, the semi-Markov approximation \cite{Jeon:2014} assumes that the times to return to visited sites are i.i.d., in contradiction with Ref.~\cite{Regnier:2023b}. In \ref{fig:St_marg}, we provide a numerical check of the logarithmic decay of Eq.~\eqref{eq:St_marg}. Including this logarithmic decay in Eq.~\eqref{eq:Nt_sum}, we obtain 
	\begin{align} \label{eq:marg_Nt}
		N(t)\propto \frac{t}{\ln t}  \; ,
	\end{align}
	so that we find, up to log-corrections, a linear growth, which is numerically verified in \ref{fig:Nt_Tn}(c).
	
	\subsection{Fluctuations of the number of visited sites}
	
	Besides $N(t)=\langle\mathcal{N}(t)\rangle$, the fluctuations of $\mathcal{N}(t)$ are irrelevant if they do not grow faster than the mean. In this paragraph, we show that indeed, for any value of $\mu$, this is the case for the number of distinct sites visited. 
	
	For the recurrent case $\mu<1$, the standard deviation of $\mathcal{N}(t)$ grows with $t$ in the same manner as its mean because of the scale-invariance of $\mathcal{N}(t)$, as we proceed to show. We first consider the first passage time $T(x)$ to a given point $x$  which obeys $T(tx) \,{\buildrel d \over =}\, t^{1/H}T(x)$ (equality in terms of distribution) \cite{levernier2019survival,Guerin:2016} using that $\mathbb{P}(T(x)\geq t)=f(x/t^{H})$ for recurrent fBm. Consequently, by noting that the number of sites visited by time $t$ is the number of sites whose first passage time to is smaller than $t$, we get that
	\begin{align}
		\frac{\mathcal{N}(t)}{t^{d H}}&=\sum_{x\in \mathbb{Z}^d} \frac{H(T(x)\leq t)}{t^{dH}}\\
		&\,{\buildrel d \over =}\, \sum_{x\in \mathbb{Z}^d} \frac{H(T(x/t^{H})\leq 1)}{t^{d H}}\\
		&\sim \int \frac{{\rm d}^dx}{t^{d H}}H(T(x/t^{H})\leq 1)\\
		&=\int {\rm d}^dx'H(T(x')\leq 1) \label{eq:sc_inv_rec_int}
	\end{align}
	where $H(\ldots)$ is the indicator function of the event in the parentheses and the last line is a random variable independent of $t$. 
	This scale invariance of $\mathcal{N}(t)$ implies the one of $T_n=\sum_{k=0}^{n-1}\tau_k$ the time to visit $n$ sites  ($T_n/n^{1/d H}$ does not depend on $n$) as 
	\begin{align}
		\frac{1}{n^{1/dH}}T_n&=\frac{1}{n^{1/d H}}\inf \lbrace t|\mathcal{N}(t)=n\rbrace  \\
		&\,{\buildrel d \over =}\,\frac{1}{n^{1/d H}}\inf \lbrace t|\mathcal{N}(t/n^{1/d H})=1\rbrace  \\
		&= \inf \lbrace t'|\mathcal{N}(t')=1\rbrace \; \label{eq:Tn_scInv}
	\end{align}
	where the last line does not contain any $n$ dependence. 
	
	For the marginal and transient fBm ($\mu\geq 1$), $ \mathcal{N}(t)$ is not scale-invariant. However, we can show that the fluctuations are still at most of the order of magnitude of the mean. Because the number of distinct sites visited is a sum of $t$ random variables equal to $0$ or $1$, the standard deviation of $\mathcal{N}(t)$ is at most $t$ (up to log corrections in the marginal case). Indeed,
	\begin{align}
		\sqrt{\text{Var}\left[\mathcal{N}(t)\right]}&=\sqrt{\sum_{0\leq t',t''\leq t}\text{Cov}\left[\Delta_{t'},\Delta_{t''}\right]}\\
		&\leq \sqrt{\sum_{0\leq t',t''\leq t} \sqrt{\text{Var}\left[\Delta_{t'}\right]\text{Var}\left[\Delta_{t''}\right]}}\\
		&=\sum_{0\leq t' \leq t}\sqrt{\text{Var}\left[\Delta_{t'}\right]}\\
		&=\sum_{0\leq t' \leq t} \sqrt{\left\langle\Delta_{t'}\right\rangle \left(1- \left\langle\Delta_{t'}\right\rangle\right)}=O(t)=O(N(t))
	\end{align}
	Thus, fluctuations of $\mathcal{N}(t)$ are (at most) of the same order of magnitude of the mean whatever the value of $\mu$: one can take $\mathcal{N}(t)=n\sim N(t)  \sim t^{\min(1,\mu)}$ (up to log corrections) in the sense that the order of magnitude of $\mathcal{N}(t)$ is given by its mean.
	\\
	
	\subsection{Properties of the visited domain: largest fully visited ball}
\begin{figure}[th!]
		\centering
		\includegraphics[width=\columnwidth]{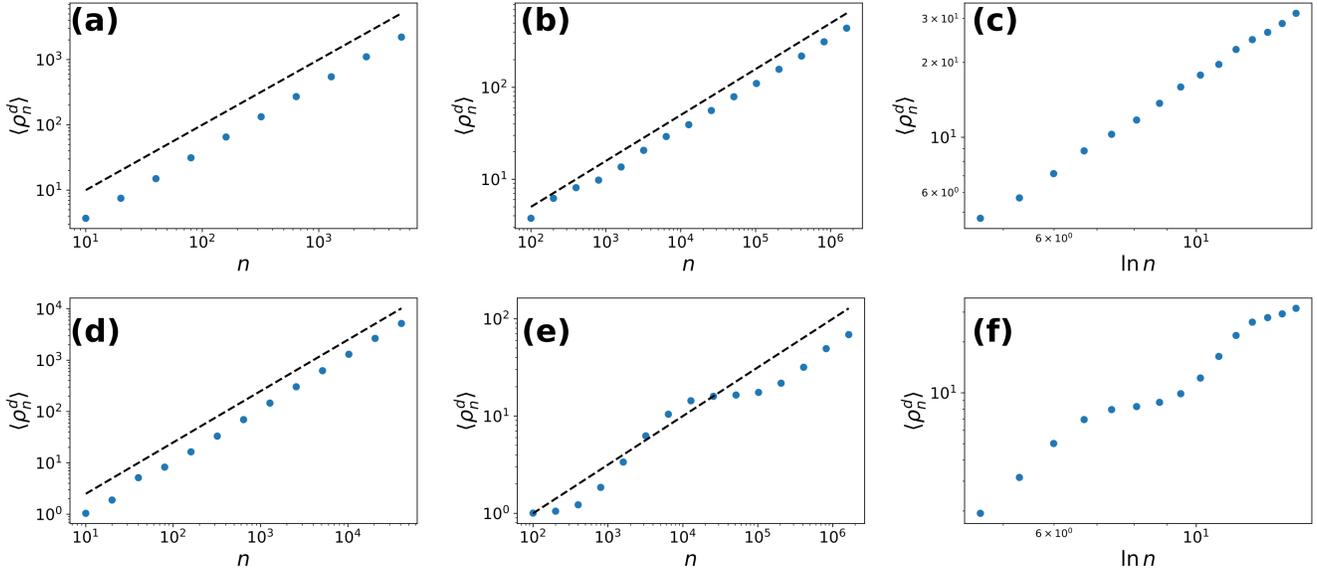}
		\caption{{\bf Typical volume of the largest ball fully visited ball.} (a) $2d$ fBm of $H=0.25$ (recurrent) (b) $3d$ fBm of $H=1/3$ (marginal) (c) $2d$ fBm $H=0.75$ (transient) (d) $3d$ fBm of $H=0.2$ (recurrent) (e) $4d$ fBm of $H=1/4$ (marginal) (f) $3d$ fBm of $H=0.4$ (transient). The black dashed lines stand for the prediction of Eq.~\eqref{eq:rho_n}. 
			\label{fig:rho_n}}
	\end{figure}

	Finally, let us consider another observable which is crucial in the understanding the visitation dynamics: the size of the largest fully visited domain. For a given number of sites visited $n$, it was shown in \cite{Dembo:2007,Regnier:2023a,Regnier:2024} that, for Markovian RWs, the average volume of the largest fully visited ball contained within the visited domain grows as (up to log corrections)
	\begin{align}
		\label{eq:rho_n}
		\left\langle \rho_n^d \right\rangle \sim \begin{cases}
			n \; \text{(recurrent)}\\
			n^{1/2} \; \text{(marginal)}\\
			1 \; \text{(transient)\;.}
		\end{cases}
	\end{align}
	These scalings were obtained for Markovian processes, however their derivation depend only on the   recurrence or transience of the walk (given that the splitting probability keeps its scaling behavior for non-Markovian processes \cite{Levernier:2018}). In the particular case of marginal fBm $H=1/d$, the scaling observed for Markovian RWs (for which the splitting probability is typically logarithmic \cite{Redner:2001}, as indicated by the marginally recurrent property of these RWs in Sec.~\ref{sec:Nt_av}) is verified numerically in \ref{fig:rho_n}. Also,  \ref{fig:rho_n} shows the linear growth with $n$ of $\left\langle \rho_n^d \right\rangle$ in the recurrent  case along with the slow growth of $\left\langle \rho_n^d \right\rangle$ (slower than any power law in $n$) in the transient case, in line with Eq.~\eqref{eq:rho_n}.
	
\begin{figure}
		\centering
		\includegraphics[width=\columnwidth]{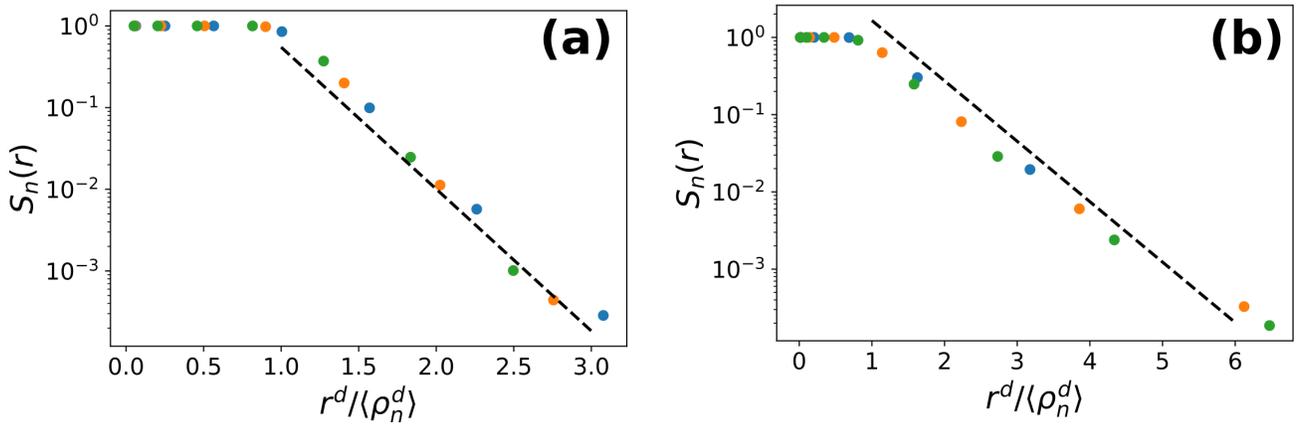}
		\caption{{\bf Distribution of the radius of the largest fully visited ball.} The distribution is shown for (a) $2d$ fBm of $H=0.75$ (transient) (b) $3d$ fBm of $H=0.33$ (marginal), for $n=100\times 2^k$, $k=7$, $8$ and $9$ (increasing values of $n$ are represented in blue, orange and green). Dashed lines represent an exponential decay.}
		\label{fig:Qn}
	\end{figure}

	Next, we check the functional form of $Q_n(r)$, the distribution of the radius of the largest ball fully visited when $n$ sites have been visited, involved in Eq.~(9) of the main text related to the stretched exponential regime. For practical reasons we consider the integral of $Q_n(r)$,
	\begin{align}
		S_n(r)=\int_r^\infty Q_n(r'){\rm d}r' =\exp\left[-r^d/ \left\langle \rho_n^d\right\rangle  \right].\label{eq:S_nr}
	\end{align}
	This quantity can be checked numerically more precisely than $Q_n(r)$. Moreover, $S_n(r)$ is the probability of having a ball of radius at least $r$ fully visited inside the visited domain and it has a simple exponential form (no eventual algebraic prefactor), as we show in~\ref{fig:Qn}. In Eq.~\eqref{eq:S_nr}, the scaling with $r^{d}$ states that, above the scale $\rho_n$, the visitation of two sites become uncorrelated, and we get back to a Poissonian distribution of the visited sites as is the case in the Rosenstock trapping problem \cite{Rosenstock:1970,Donsker:1979}. 
	
	Finally, from this volume $\left\langle \rho_n^d \right\rangle$, we define the typical time $\vartheta_n$ to escape this largest fully visited domain as (see Eq.~\eqref{eq:rho_n}):
	
	\begin{align} \label{eq:tn}
		\vartheta_n \equiv \left\langle \rho_n^d \right\rangle^{d_\text{w}/d} \propto 
		\begin{cases}
			n^{1/\mu} & \text{(recurrent)} \\
			n^{1/2} & \text{(marginal)} \\
			1 & \text{(transient)}.
		\end{cases}
	\end{align}
	
	Physically speaking, this time separates time scales at which a random walk sees the visited domain as infinite (at times much smaller than $\vartheta_n$) from times when finite-size effects start to appear. From now on, we will abridge $\left\langle \rho_n^d \right\rangle^{1/d}$ to $\rho_n$, which is to be interpreted as the typical correlation length of visited domains.
	
	\section{Alternative (mean-field) approach to the exponent characterizing recurrent visitation}
	
	In this section, we show that the exponent of the algebraic decay presented in Eq.~(8) of the main text can be derived also by using a mean-field approach.
	
	For recurrent fBm, we have shown in Eq.~\eqref{eq:Tn_scInv} that  $T_n/n^{1/\mu}$ is independent of $n$. Consequently, using that $T_n=\sum_{k=0}^{n-1} \tau_k$, we deduce that
	\begin{align}
		\left\langle \tau_n \right\rangle =\left\langle T_{n+1} \right\rangle-
		\left\langle T_n \right\rangle \propto n^{1/\mu-1} \label{eq:avg_taun}
	\end{align}
	Then, by viewing $\tau_n$ as the exit time from an infinite domain at times smaller than the typical time $\vartheta_n=n^{1/\mu}$ (see Eq.~\eqref{eq:tn}), $\tau_n$ has an algebraic distribution independent of $n$, whose decay is of unknown exponent we note $1+\delta$.
	The average value of $\tau_n$ is then governed by the early-time regime, 
	\begin{align}
		\langle \tau_n \rangle \propto \int_1^{n^{1/\mu}} \frac{\tau {\rm d}\tau}{\tau^{1+\delta}}\propto n^{1/\mu-\delta/\mu}
	\end{align}
	which imposes $\delta=\mu$, see Eq.~\eqref{eq:avg_taun}. 
	
	In the case of marginal fBm, the same argument applies, using that the visitation rate decays as $1/\ln t$ (see  Eq.~\eqref{eq:St_marg}). Thus we expect $\left\langle T_n \right\rangle \sim n \ln n $ (inverting the relation $n\sim t/\ln t$, Eq.~\eqref{eq:marg_Nt}). Finally, using the typical time to escape the semi-infinite domain $\vartheta_n=\rho_n^{1/H} =n^{1/2}$ (see Eq.~\eqref{eq:tn}), the average inter-visit time grows logarithmically with $n$ and the exponent $\delta=\mu=1$.\\
	
	\section{Effective independence criterion of inter-visit times}
	In this section, we justify the effective independence criterion between the inter-visit times $\tau_n$ used in Eq.~(6) of the main text in the recurrent case. For the marginal case, we expect that our physical interpretation of decorrelation of the inter-visit times still holds, as checked numerically in the next section.\\
	
	To start, we note that, due to the scale-invariant property of the number of distinct sites visited derived in Sec.~\ref{sec:Prop} in the recurrent case, the distribution $F_n(\tau)=\mathbb{P}(\tau_n=\tau)$ of the inter-visit time $\tau_n$ should only depend on  three time scales: $\tau$, $ n^{1/\mu}$ the typical time to visit $n$ sites and $\Delta n^{1/\mu}=1$ the typical time to exit the unit hypercube. Thus, by dimensional analysis and because we consider times $\tau,n^{1/\mu}\gg \Delta n^{1/\mu}=1$, the distribution should have the following general functional form
	\begin{align}
		F_n(\tau)=\left(\frac{\Delta n}{n}\right)^{\gamma}\frac{1}{\tau}\psi\left(\tau/n^{1/\mu} \right) \; . \label{eq:scaling_rec}
	\end{align}
	To determine the exponent $\gamma$, we use that $\left\langle T_n \right\rangle =\sum_{k=0}^{n-1} \left\langle \tau_k \right\rangle \propto n^{1/\mu}$ so that 
	\begin{align}
		n^{1/\mu-1}\propto \left\langle \tau_n\right\rangle =\int_{1}^\infty {\rm d}\tau \tau \left(\frac{\Delta n}{n}\right)^{\gamma}\frac{1}{\tau}\psi\left(\tau/n^{1/\mu} \right) \propto n^{1/\mu-\gamma} \; ,
	\end{align}
	which implies $\gamma=1$. 
	
	From this, we deduce the scaling with $n$ of the standard deviation of $\tau_n$. Indeed, using Eq.~\eqref{eq:scaling_rec} to compute the second moment of $\tau_n$ we get that
	\begin{align}
		\left\langle \tau_n ^2 \right\rangle =\int_{1}^\infty {\rm d}\tau  \tau^2\left(\frac{\Delta n}{n}\right)\frac{1}{\tau}\psi\left(\tau/n^{1/\mu} \right) \propto n^{2/\mu-1} \gg \left\langle \tau_n  \right\rangle^2.
	\end{align}
	Finally,
	\begin{align}
		\sqrt{\text{Var}\left[ \tau_n \right]}\propto \sqrt{\left\langle  \tau_n^2 \right\rangle } \propto n^{1/\mu-1/2} \; . \label{eq:true_var_tau}
	\end{align}
	\\
	
	We now propose an argument showing that the correlations between inter-visit times are not relevant to the determination of the law of their maximum. For this, we apply the argument of \cite{Carpentier:2001} (also used for the characterization of the extreme values of the inter-visit times \cite{Regnier:2024} of Markovian RWs and record ages \cite{Regnier:2023a} of 1d stochastic processes), which compares the typical fluctuations of $\max\limits_{n\leq k<n+m} \tau_k$ obtained by supposing that the $\tau_k$ are independent to the typical correlations between the $\tau_k$.  Based on Eqs.~(4) and (5) of the main text, it follows for $1\ll m\ll n$ that $\max\limits_{n\leq k<n+m} \tau_k \sim m^{1/\mu}$, implying that the fluctuations of  $\max\limits_{n\leq k<n+m} \tau_k$ increase with $m$, see Eq.~\eqref{eq:true_var_tau}. However, we can show that the typical correlations between the inter-visit times decrease with $m$ in the limit  $1\ll m\ll n$. 
	
	To do so, we start from the scaling with $n$ of $T_n$ fluctuations:
	\begin{align}
		\text{Var}\left[ T_n \right]=\sum_{1 \leq k,l\leq n} \text{Cov}\left[ \tau_k,\tau_l \right] \propto n^{2/\mu}
	\end{align}
	If we suppose that the covariance is not decreasing i.e. $\text{Cov}\left[ \tau_k,\tau_l \right] \geq \text{Var}(\tau_k)$ if $k<l$, then we end up with 
	\begin{align}
		n^{2/\mu}\propto \text{Var}\left[ T_n \right] \geq  \sum_{1 \leq k\leq n} \text{Var}\left[ \tau_k \right] (2n-2k+1)  
	\end{align}
	which would finally lead to 
	\begin{align}
		\sqrt{\text{Var}\left[ \tau_n \right]}  =O( n^{1/\mu-1})\label{eq:false_var}
	\end{align}
	However, this upper bound scaling with $n$ enters in contradiction with the scaling obtained in Eq.~\eqref{eq:true_var_tau}. It implies that the correlation between the inter-visit times should decay.  
	
	Consequently, the fluctuations of the maximum $\max\limits_{n\leq k<n+m} \tau_k$ grow with $m$ while the correlations between $\tau_k$ decay. This decay should be algebraic (to make Eq. \eqref{eq:false_var} compatible with Eq.~\eqref{eq:true_var_tau}). This means that one can neglect the correlations between $\tau_k$ for determining the maximum's law, as discussed in the main text. This result is in line with that of the mathematical work of Ref.~ \cite{Samorodnitsky:2004} (see also Ref. \cite{Schumann:2012}, where a similar argument can be found for analysis of heavy-tail correlated time series).\\

	\section{Numerical check of the effective independence of inter-visit times}
	In this section, we provide a numerical check of the effective independence of the intervisit times by computing the distribution of the maximum with and without independence.\\
	
\begin{figure}[th!]
		\centering
		\includegraphics[width=\columnwidth]{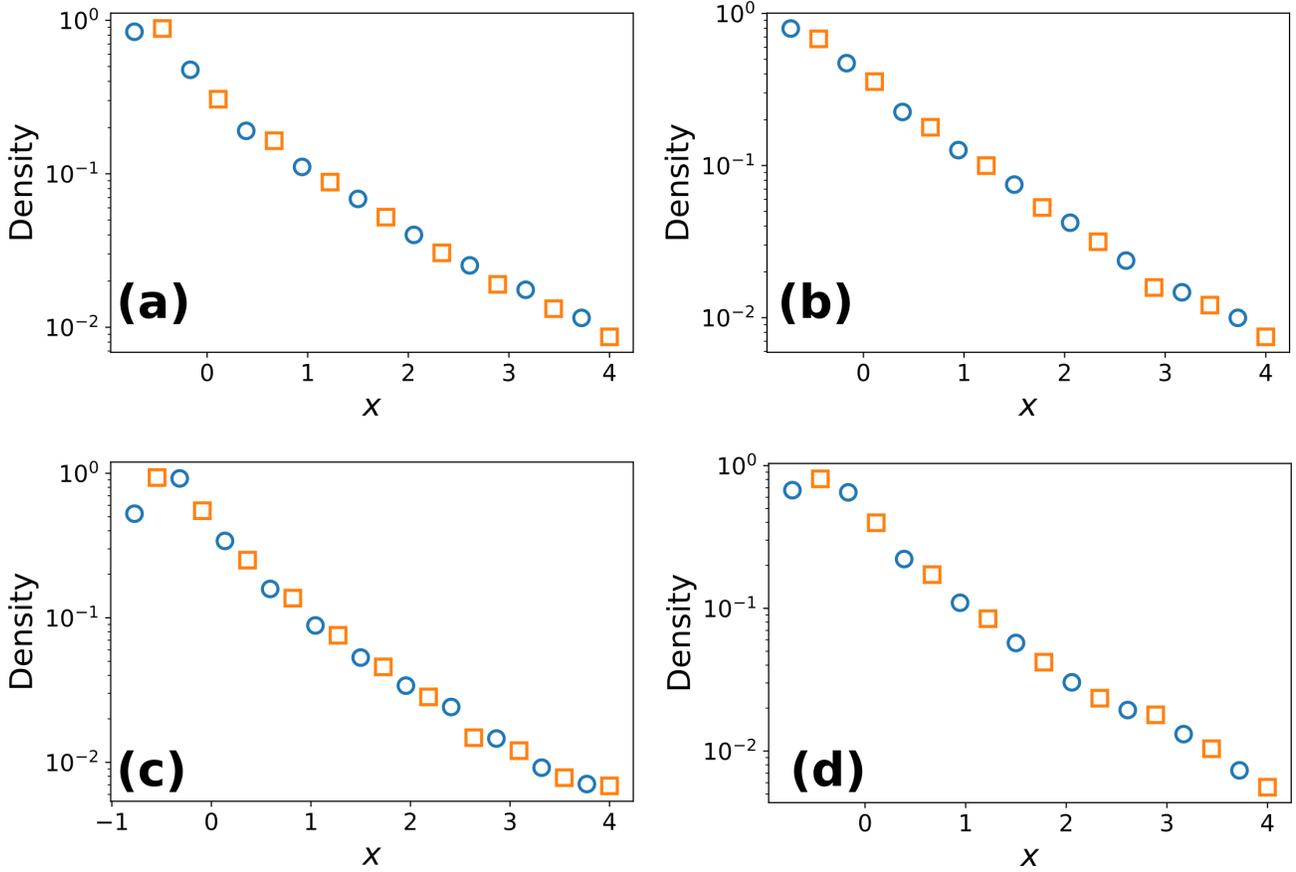}
		\caption{{\bf Comparison of the maximum statistics with and without independence of inter-visit times.} Distribution of the centered and normalized maximum, $x \equiv \left(X-\left\langle X\right\rangle\right)/\sqrt{\text{Var}\left[X \right]}$, where $X=\max\limits_{n\leq k<n+m} \tau_k$ of the $\tau_k$ being drawn from a given trajectory (blue circles)  or from independent trajectories for every $k$ (orange squares) for fBm in {\bf (a)} $2d$, $H=0.25$ ($n=800$, $m=40$), {\bf (b)} $3d$, $H=0.2$ ($n=800$, $m=40$), {\bf (c)} $3d$, $H=1/3$ ($n=3200$, $m=40$), {\bf (d)}  $4d$, $H=0.25$ ($n=3200$, $m=40$).}
		\label{fig:Tnm}
	\end{figure}

	In \ref{fig:Tnm}, we compare the distribution of the maximum of $m$ random variables $\max\limits_{n\leq k<n+m} \tau_k$ with the distribution assuming that the $m$ random variables are independent. To obtain the distribution with independent $\tau_k$, we first generate a large number of fBm paths. For each path, we compute the list of $\tau_k$ values. To construct a realization of $\max\limits_{n \leq k < n+m} \tau_k$ with independent $\tau_k$, we proceed as follows: for each $k$ in the range from $n$ to $n+m$, we randomly select one of the generated fBm paths and use the corresponding $\tau_k$ from that path. This gives us $m$ independent $\tau_k$ values. Finally, we compute the maximum of these $m$ random variables.\\
	As can be seen in the figure, the two distributions of the maximum with and without independence of the intervisit times are practically identical, which further emphasizes that correlations can be neglected when one is interested in the maximum asymptotic distribution, and confirming the validity of the approximation made in Eq.~(6) of the main text.
	
	\newpage
	
	\section{Numerical method for the long-time statistics}
	
	In this section, we detail the simulation methods (consisting from several algorithms) presented in the main text, which allow us to sample the high inter-visit times or extremely low likelihood values for fBm. It was developed by Hartmann et al. \cite{Hartmann:2013,Hartmann:2024} and we adapted it for our purposes.
	
	\subsection{The Davies-Harte Algorithm: Circulant Embedding Method}
	
	The Davies-Harte algorithm, also known as the circulant embedding method \cite{Davies:1987,Wood:1994}, is a technique used to generate a time series of correlated increments for a fractional Brownian motion (fBm) of Hurst index $H$. This method starts by generating $2K$ independent Gaussian variables, each with a mean of zero and a standard deviation of one, denoted as $\boldsymbol{\xi}=(\xi_0,\ldots,\xi_{2K-1})$. These variables are then used to produce $K$ correlated increments $\boldsymbol{\eta}=(\eta_0,\ldots,\eta_{K-1})$.
	
	The process begins by defining the correlation function $C(m)$ of the increments, which is expected from the formula Eq.~(1) of the main text. This correlation function is defined as (independently of $l$):
	\begin{align} \label{eq:cov_steps}
		C(m)=\left\langle \eta_l \eta_{l+m} \right\rangle=D\left(|m+1|^{2H}+|m-1|^{2H}-2|m|^{2H}\right)
	\end{align}
	where $D$ is such that $\sqrt{C(0)}=1/2$ (the typical step length of the random walk is half the lattice spacing).
	Next, a function $\mathcal{C}(m)$ is defined for $m\in[0,2K-1]$, which is derived from $C(m)$ as:
	\begin{align}
		\mathcal{C}(m)=C(\min(m,2K-m))
	\end{align}
	This function is then fast Fourier transformed, enabling the algorithm's complexity to remain linear with respect to the system size $K$. The transformed coefficients are given by:
	\begin{align}
		\hat{c}_k\equiv \sum_{m=0}^{2K-1}\mathcal{C}(m)e^{-i \pi k m/K} \; .
	\end{align}
	Subsequently, the increments $\boldsymbol{\eta}$ are obtained by summing the real and imaginary parts of the inverse Fourier transform of the $2K$ coefficients:
	\begin{align}
		\eta_m=\frac{1}{2K}\left[ \text{Re}\left(\sum_{k=0}^{2K-1}\sqrt{2K \hat{c}_k} \xi_k e^{i\pi k m/K}\right)+\text{Im}\left(\sum_{k=0}^{2K-1}\sqrt{2K \hat{c}_k} \xi_k e^{i\pi k m/K}\right)\right]
	\end{align}
	These increments exhibit the expected correlation function \eqref{eq:cov_steps} and follow a Gaussian distribution.
	
	By summing these increments, a trajectory $\boldsymbol{x}=(x_0,\ldots,x_{K-1})$ of an fBm is obtained:
	\begin{align}
		x_t=\sum_{k=0}^{t-1} \eta_k \; .
	\end{align}
	To generate a trajectory in dimension $d$, this algorithm must be executed $d$ times independently.\\
	
	To derive the inter-visit times from the increments $\boldsymbol{\eta}$, the $d$-dimensional domain is discretized into hypercubic unit cells. A new site is considered visited whenever the sampled trajectory enters an unvisited unit cell. This yields an algorithm that computes an intervisit time, denoted as $\tau=\tau(\boldsymbol{\xi})$, where vector $\boldsymbol{\xi}$ consists from $d$ fBm trajectories each of $2K$ points. We stress that the attribution of the inter-visit times to $\boldsymbol{\xi}$ is more practical, since $\boldsymbol{\xi}$ consists from independent variables.

	\subsection{Importance sampling}
	
	We are interested in sampling of large inter-visit times and obtaining their distribution $F_n(\tau)$. In the following, we forget the index $n$ which does not play any role in the algorithm except for the choice of a high enough $K$-value. We stress that the preceding algorithm for the generation of fBm paths is not powerful enough for the  observation of the stretched-exponential and exponential regimes at long times, see Eq.~(9) in the main text.
	Thus, we couple this algorithm with a Monte Carlo Markov chain approach following the Metropolis-Hasting prescription \cite{Metropolis:1953,Tokdar:2010}. \\
	
	First, we have to choose a bias noted $g(\tau(\boldsymbol{\xi}))$ which depends on the configuration $\boldsymbol{\xi}$. Here, we must choose a bias which grows with increasing intervisit time $\tau(\boldsymbol{\xi})$, such that the reweighted probability 
	\begin{align}
		F_g(\tau) \equiv \frac{F(\tau)g(\tau)}{\sum_\tau F(\tau)g(\tau)}\label{eq:bias_dis}
	\end{align}
	selects high $\tau$ values. Because the normalization constant is unknown, one needs to perform a Monte Carlo Markov Chain (MC-MC) algorithm to sample $F_g$. To do so, we use the standard Metropolis-Hasting prescription \cite{Metropolis:1953,Tokdar:2010}.
	
	From a configuration $\boldsymbol{\xi}$ we draw a new configuration $\boldsymbol{\xi}'$ by randomly choosing components of $\boldsymbol{\xi}$ and redrawing Gaussian variables of mean $0$ and standard deviation $1/2$ for them. From this new configuration $\boldsymbol{\xi}'$ we obtain a new intervisit time $\tau'=\tau(\boldsymbol{\xi}')$. Then, we accept the change with a probability
	\begin{align}
		\min\left(1, \frac{g(\tau')}{g(\tau)} \right) \; .
	\end{align}
	Once we have drawn the inter-visit times  using this algorithm, we obtain the true inter-visit time statistics as
	\begin{align}
		F(\tau)\propto F_g(\tau)g(\tau)^{-1}
	\end{align}
	where the determination of the normalizing constant is explained in the following part.
	
	\subsection{Parallel tempering} 
	
	As highlighted in the previous part, the choice of the bias $g(\tau)$ is crucial when it comes to sampling of different likelihoods of the inter-visit times. However, choosing one bias does not always allow to sample the whole domain of interest. Additionally, if the bias favours very unlikely inter-visit times, the algorithm might take a very long time to equilibrate to the statistics of $F_g$. This is why one often uses parallel tempering \cite{Earl:2005,Hartmann:2024}: the bias $g_\beta(\tau)$ is indexed by an inverse temperature $\beta$, such that higher $\beta$ leads to biased inter-visit time statistics $F_{g_\beta}=F_\beta$ centered at higher values of $\tau$. We perform in parallel the MC-MC algorithm described above for different inverse temperatures. Every ten steps, we choose two consecutive inverse temperatures $\beta$ and $\beta'$ associated with configurations $\boldsymbol{\xi}$ and $\boldsymbol{\xi}'$ and hence with inter-visit times $\tau$ and $\tau'$, respectively. Then, we exchange the two configurations with probability given by
	\begin{align}
		\min\left(1,\frac{g_\beta(\tau')g_{\beta'}(\tau)}{g_\beta(\tau)g_{\beta'}(\tau')} \right) \; .
	\end{align}
	Once we sampled every biased inter-visit time statistics $F_\beta$ (up to the normalization constant), we then put them together by jointly normalizing them: we start from low inverse temperature $\beta$, which we normalize as
	\begin{align}
		F(\tau)= F_\beta(\tau)g_\beta(\tau)^{-1}/\left(\sum_{\tau'} F_\beta(\tau')g_\beta(\tau')^{-1} \right),
	\end{align}
	then, we let the higher $\beta$ values of the shifted biased inter-visit time statistics $F_\beta(\tau)g_\beta(\tau)^{-1}$ coincide with the lowest ones successively.\\

	We are left with the choice of the family of functions $\lbrace g_\beta \rbrace$ which obey the previous requisites (grows with $\tau$ and samples typically larger inter-visit times with higher $\beta$).
	
	\subsection{Choice of the importance sampling weight}
	For the bias $g_\beta$, we make the standard choice of an exponential function of the inter-visit time to a power of $\kappa$ (that we determine below) times the inverse temperature $\beta$,
	\begin{align}
		g_\beta(\tau)=\exp\left[-\beta \tau^{\kappa} \right].
	\end{align}
	With this choice of $g_\beta$, the biased distribution can be expressed as (using Eq.~\eqref{eq:bias_dis})
	\begin{align}
		F_\beta(\tau)\propto F(\tau) \exp\left[-\beta\tau^{\kappa} \right].
	\end{align}
	In order to use the importance sampling algorithm to sample large values of the inter-visit time $\tau$, one has to choose the exponent $\kappa$ in such a way that the probability distribution $F_\beta$ is well-defined (which is always the case if $\beta>0$ as the exponential is smaller than one),  and second, that changing the value of $\beta$ allows to sample effectively different typical values of $\tau$. 
	
	To proceed, we recall first that we expect $F(\tau)$ to be stretched exponentially decaying with $\tau$, $F(\tau)\sim \exp\left[-\tau^{\mu/(1+\mu)} \right]$. Second, we find the typical value $\tau^*(\beta)$ by minimizing the argument of the negative logarithmic likelihood at large $\tau$, $-\ln F_\beta(\tau)\approx \beta\tau^{\kappa}+\tau^{\mu/(1+\mu)}$. This results in 
	\begin{align}
		\tau^*(\beta)\propto \left(-\frac{\kappa \beta(1+\mu)}{\mu} \right)^{1/(\mu/(1+\mu)-\kappa)}>0
	\end{align}
	Thus, we observe that the minimum exists if and only if $\kappa \beta <0$. We notice that by taking $\kappa=\mu/(1+\mu)-1<0$ and $\beta>0$, the typical value of $\tau$ grows linearly with $\beta$. We choose hence $\kappa=-1/(1+\mu)$ to facilitate our numerical procedure.
	
	\subsection{Choice of parameters}
	
	For every number of sites visited $n$ and type of RW (except the low values $n=10$, $n=20$ and $n=25$), we took $K=30 n $ and checked that indeed no trajectory leading to $n+1$ visited sites longer than $K$ occurred during the $10^7$ MC-MC steps performed.  We also took $24$ different values of the inverse temperature $\beta \in [4, 40 000]$ equally spaced on a logarithmic scale. For the values $n=10$, $n=20$ and $n=25$, we took $K=4000 $ and $30$ different values of the inverse temperature $\beta \in [4, 4 000 000]$ equally spaced on a logarithmic scale.
	
	\section{Data analysis}
	
	In this section, we describe the numerical method used to analyze the experimental datasets shown in Fig.~4 of the main text.\\
	
	For each data set, we have a given number of $2d$ trajectories, of length $2048$ time steps, $(x_1(t),x_2(t))_{t \leq 2048}$. Space is discretized, see Fig.~1 of the main text, with a lattice spacing of $2\sigma$, where $\sigma$ is the standard deviation of the steps, $x_i(t+1)-x_i(t)$. This choice of $2\sigma$ is such that the RW almost never crosses two sites at a time even when close to the boundary of a site of the grid, while still making steps large enough to visit a new site in a small number of jumps. Then, we obtain the distributions of the inter-visit times by splitting each of the $2d$ trajectory into $50$ trajectories of equal length, as displayed in Fig.~4 of the main text.